\documentclass[12pt]{iopart}

\usepackage{iopams}
\usepackage{graphicx}

\begin{document}

\title[Are High-$p_\mathrm{T}$ Pions Suppressed in Pb+Pb Collisions at
$\sqrt{s_\mathrm{NN}} = 17.3$\,GeV?]{\boldmath Are High-$p_\mathrm{T}$
  Pions Suppressed in Pb+Pb Collisions at $\sqrt{s_\mathrm{NN}} =
  17.3$\,GeV? \unboldmath}

\author{K. Reygers for the WA98\footnote{For the full list of WA98
    authors, see appendix 'Collaborations' of this volume.}
  collaboration}

\address{University of M{\"u}nster, Institut f{\"u}r Kernphysik,
Wilhelm-Klemm-Stra{\ss}e 9,\\ 48149 M{\"u}nster, Germany}
\ead{reygers@uni-muenster.de}
\begin{abstract}
  Transverse momentum spectra of neutral pions in the range 
  $0.7<p_\mathrm{T}<3.2$\,GeV/$c$ have been measured at $2.3 \lesssim
  \eta \lesssim 3.0$ by the WA98 experiment in p+C and p+Pb collisions
  at $\sqrt{s_\mathrm{NN}} = 17.3$\,GeV.  Scaled by the number of
  nucleon-nucleon collisions ($N_\mathrm{coll}$) the $\pi^0$ yields in
  p+C and p+Pb at $p_\mathrm{T} \approx 2.0 - 2.5$\,GeV/$c$ are higher
  than the respective yields in central Pb+Pb collisions with
  $N_\mathrm{part} \gtrsim 300$. This observation is qualitatively
  consistent with expectations from parton energy loss.
\end{abstract}


\section{Introduction}
The suppression of high-$p_\mathrm{T}$ hadrons observed at RHIC in
central Cu+Cu and Au+Au collisions at $\sqrt{s_\mathrm{NN}} =
200$\,GeV is most naturally explained by jet-quenching models which
attribute the suppression to parton energy loss in a quark-gluon
plasma (QGP).  With the aid of these models the strength of the
suppression can be related to medium properties such as the initial
gluon density.  The dependence of the hadron suppression on the
transverse momentum ($p_\mathrm{T}$) of the produced particles and on
the centrality of the collisions was studied at RHIC. However, not
much is known about the dependence on the center-of-mass energy
($\sqrt{s_\mathrm{NN}}$) of the collision. In particular, it is an
open question whether jet-quenching plays a role in Pb+Pb collisions
at the CERN SPS energy of $\sqrt{s_\mathrm{NN}} = 17.3$\,GeV
\cite{d'Enterria:2004ig,Blume:2006va}.

In central Pb+Pb collisions at $\sqrt{s_\mathrm{NN}} = 17.3$\,GeV the
initial energy density as estimated from the measured transverse
energy is above the critical value $\varepsilon_\mathrm{c} \approx
0.7$\,GeV/fm$^3$ for the transition to the QGP \cite{Aggarwal:2000bc}.
Thus, it is reasonable to expect that high-$p_\mathrm{T}$ particle
production is affected by the created medium. The problem at the CERN
SPS energy is that p+p reference data are not available. Therefore,
different p+p parameterizations have been employed to study nuclear
effects in Pb+Pb collisions \cite{d'Enterria:2004ig,Aggarwal:2001gn}.
Moreover, hadron suppression due to parton energy loss might be
compensated by an enhancement due to initial state multiple soft
scattering of the incoming partons (``nuclear
$k_\mathrm{T}$-enhancement''), an effect which is expected to be
stronger at $\sqrt{s_\mathrm{NN}} = 17.3$\,GeV than at
$\sqrt{s_\mathrm{NN}} = 200$\,GeV
\cite{Blume:2006va,Barnafoldi:2003kb}.

The WA98 collaboration has measured the centrality dependence of
neutral pion production in Pb+Pb collisions at $\sqrt{s_\mathrm{NN}} =
17.3$\,GeV in the range $0.5 \lesssim p_\mathrm{T} \lesssim
4$\,GeV/$c$ \cite{Aggarwal:2001gn}.  Here, we present data on neutral
pion production in p+C and p+Pb collisions at the same energy which
were taken in 1996. The p+C $\pi^0$ spectrum can be used as a
replacement for a p+p reference since the nuclear
$k_\mathrm{T}$-enhancement is expected to be small
\cite{Barnafoldi:2003kb}. By comparing the $\pi^0$ production in p+C
and p+Pb collisions one can study the strength of the nuclear
$k_\mathrm{T}$-enhancement at the CERN SPS energy. Finally, by using
the p+Pb spectrum as a reference for the Pb+Pb data it is expected
that the nuclear $k_\mathrm{T}$-enhancement in Pb+Pb is partially
cancelled by the enhancement in p+Pb so that possible effects of
parton energy loss can be seen more clearly.

\section{Results}
In the WA98 experiment $\pi^0$ yields were measured by detecting
photons from the decay $\pi^0 \rightarrow \gamma\gamma$ with a
highly-segmented lead glass calorimeter. This detector was located
21.5\,m downstream of the target and subtended the pseudorapidity
range $2.3 \lesssim \eta \lesssim 3.0$. The measurement of the
transverse energy ($E_\mathrm{T}$) with a hadronic calorimeter in the
range $3.5 \lesssim \eta \lesssim 5.5$ provided the minimum bias
trigger. The measured minimum bias cross section $\sigma_\mathrm{mb}$
for p+C (p+Pb) of 170\,mb (1341\,mb) corresponds to 74\,\% (76\,\%) of
the total geometric cross section. A $p_\mathrm{T}$ reach of the
$\pi^0$ spectra in p+C and p+Pb comparable to that in Pb+Pb could only
be achieved by employing a high-energy photon (HEP) trigger based on
the energy signal seen by the lead glass calorimeter.

\begin{figure}[t]
\includegraphics[width=\textwidth]{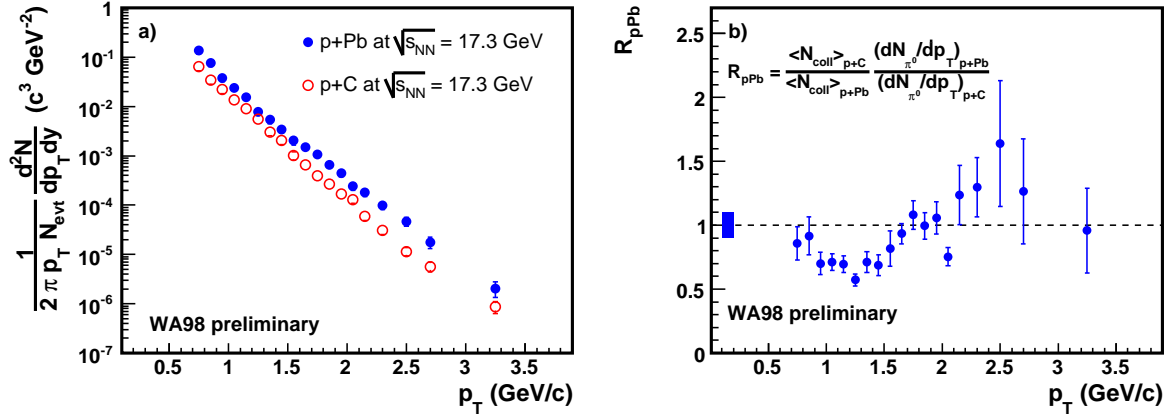}
\caption{a) Invariant neutral pion yield in minimum bias p+C and p+Pb
  collisions. b) Ratio p+Pb/p+C of the $\pi^0$ yields normalized to
  the respective $N_\mathrm{coll}$ values.}
\label{fig:spectra}
\end{figure}
The fully corrected invariant $\pi^0$ yields in p+C and p+Pb
collisions at $\sqrt{s_\mathrm{NN}} = 17.3$\,GeV are shown in
Figure~\ref{fig:spectra}a. For both spectra the transition between the
minimum bias and the HEP sample takes place at $p_\mathrm{T} =
1.7$\,GeV/$c$. Figure~\ref{fig:spectra}b shows the ratio of the
$\pi^0$ spectra where each spectrum was normalized to the number of
binary nucleon-nucleon collisions. A Glauber Monte Carlo calculation
with a nucleon-nucleon inelastic cross section of
$\sigma_\mathrm{inel}^\mathrm{NN} = 32$\,mb yields $\langle
N_\mathrm{coll} \rangle_\mathrm{p+C} = 1.7 \pm 0.2$ and $\langle
N_\mathrm{coll} \rangle_\mathrm{p+Pb} = 4.2 \pm 0.4$. The centrality
bias due to the minimum bias trigger was taken into account in the
calculation. Consistent with the expectation of a stronger nuclear
$k_\mathrm{T}$-enhancement for heavier nuclei the $\pi^0$ spectrum in
p+Pb appears to be slightly flatter than in p+C.

\begin{figure}[t]
\includegraphics[width=\textwidth]{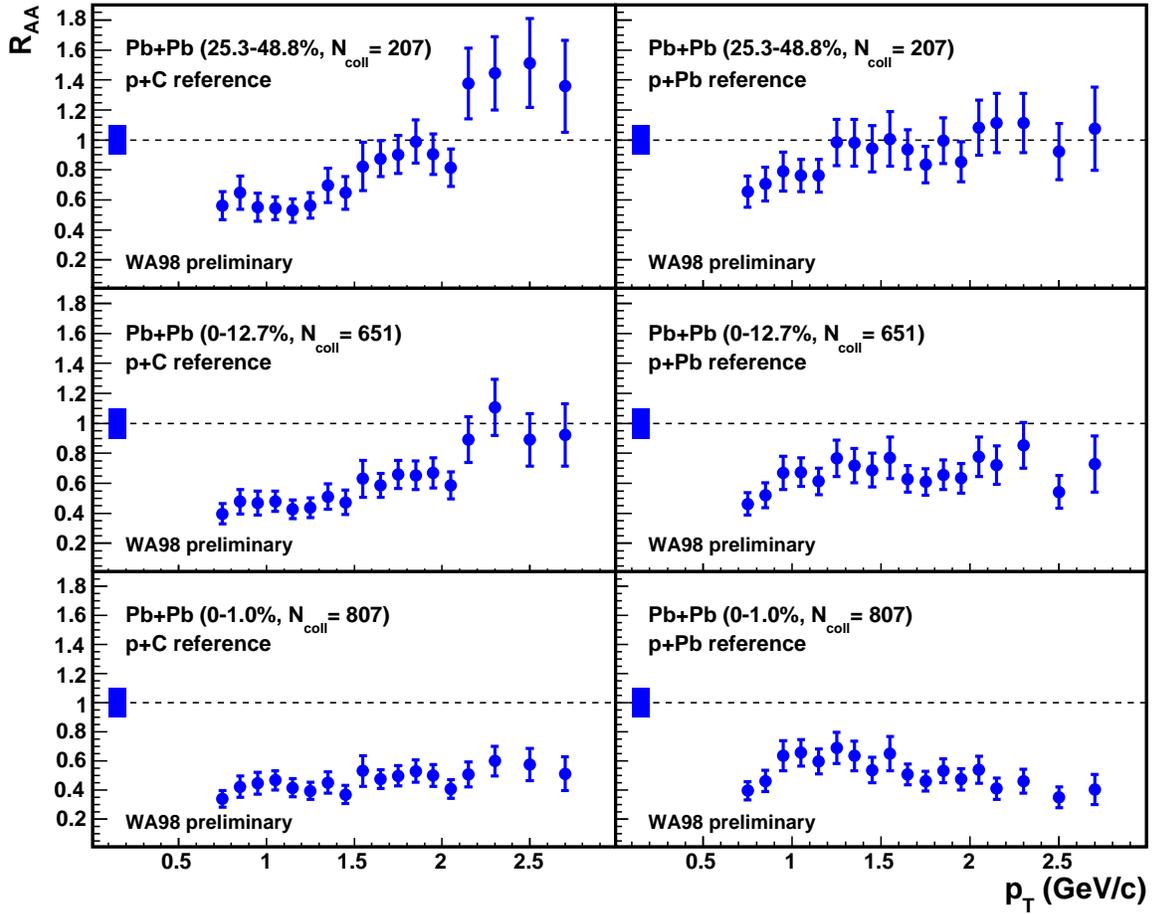}
\caption{Nuclear modification factor $R_\mathrm{AA}$ for three Pb+Pb
  centrality classes with p+C (left panel) and p+Pb (right panel) data
  as reference. The boxes indicate the uncertainty of the ratio
  $\langle T_\mathrm{p+B} \rangle / \langle T_\mathrm{Pb+Pb}
  \rangle$.}
\label{fig:raa_panel}
\end{figure}
A possible suppression of high-$p_\mathrm{T}$ $\pi^0$'s in Pb+Pb
collisions can be quantified with a nuclear modification factor
defined as
\begin{equation}
R_\mathrm{AA} = \frac{\langle T_\mathrm{p+B} \rangle}
                     {\langle T_\mathrm{A+A} \rangle} 
  \frac{\left. \mathrm{d}N_{\pi^0}/\mathrm{d}p_\mathrm{T}\right|_\mathrm{A+A}}
       {\left. \mathrm{d}N_{\pi^0}/\mathrm{d}p_\mathrm{T}\right|_\mathrm{p+B}}
\label{eq:raa}
\end{equation}
where $\langle T_\mathrm{X+Y} \rangle = \langle N_\mathrm{coll}
\rangle_\mathrm{X+Y} / \sigma_\mathrm{inel}^\mathrm{NN}$. In the
absence of nuclear effects $R_\mathrm{AA}$ is expected to be unity for
$p_\mathrm{T} \gtrsim 2$\,GeV/$c$ where hard scattering is expected to
dominate particle production.  The $\langle N_\mathrm{coll} \rangle$
values for Pb+Pb given in \cite{Aggarwal:2001gn} were determined with
the event generator VENUS by applying cuts to the simulated
$E_\mathrm{T}$ corresponding to the same fraction of
$\sigma_\mathrm{mb}^\mathrm{Pb+Pb}$ as the cuts applied to the
measured $E_\mathrm{T}$. The simulated $E_\mathrm{T}$ described the
measured $E_\mathrm{T}$ well, including the fluctuations in central
Pb+Pb collisions, \cite{Aggarwal:2000bc} so that a centrality class
corresponding to the $1\,\%$ most central collisions was defined.  By
using $\langle T_\mathrm{X+Y} \rangle$ in Eq.~\ref{eq:raa} the small
difference between $\sigma_\mathrm{inel}^\mathrm{NN}$ at
$\sqrt{s_\mathrm{NN}} = 17.3$\,GeV used in VENUS ($29.6$\,mb) and in
the Glauber Monte Carlo calculation for p+C and p+Pb ($32$\,mb) does
not affect $R_\mathrm{AA}$.

\begin{figure}[t]
\centerline{\includegraphics[width=.37\textheight]{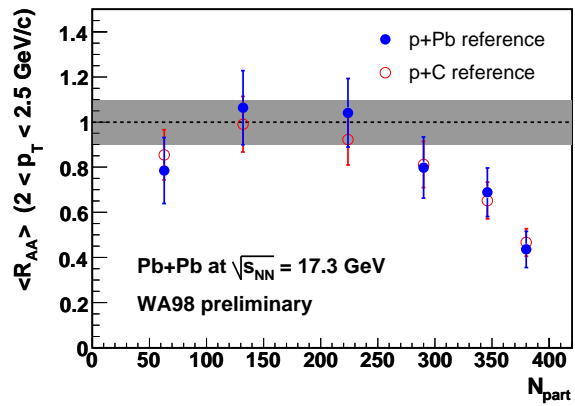}}
\caption{Centrality dependence of $\langle R_\mathrm{AA} \rangle $ for
  Pb+Pb collisions at $\sqrt{s_\mathrm{NN}} = 17.3$\,GeV. $\langle
  R_\mathrm{AA} \rangle $ was obtained by fitting
  $R_\mathrm{AA}(p_\mathrm{T})$ with a constant at $2.0 < p_\mathrm{T}
  < 2.5$\,GeV/$c$.}
\label{fig:raa_vs_npart}
\end{figure}

Figure~\ref{fig:raa_panel} shows that for semi-central Pb+Pb
collisions ($25.3-48.8$\,\% of $\sigma_\mathrm{mb}^\mathrm{Pb+Pb}$)
$R_\mathrm{AA}$ is consistent with $N_\mathrm{coll}$ scaling for
$p_\mathrm{T} \gtrsim 2$\,GeV/$c$.  A first indication of a
suppression ($R_\mathrm{AA} \approx 0.7$) is visible for the
$12.7\,\%$ most central Pb+Pb collisions using p+Pb data as reference.
For the $1\,\%$ most central Pb+Pb collisions a suppression is
observed for the p+Pb reference as well as for the p+C reference.
Moreover, a change of the shape of the $\pi^0$ spectrum for the
$1\,\%$ becomes visible. The centrality dependence of $R_\mathrm{AA}$
is shown in Figure~\ref{fig:raa_vs_npart}.  The suppression for
$N_\mathrm{part} \gtrsim 300$ appears to increase with centrality so
that the approximate $N_\mathrm{coll}$ scaling for $50 \lesssim
N_\mathrm{part} \lesssim 300$ might well be due to a compensation of
parton energy loss and nuclear $k_\mathrm{T}$-enhancement.

\section{Conclusions}
Neutral pion spectra from p+C and p+Pb collisions at
$\sqrt{s_\mathrm{NN}} = 17.3$\,GeV have been measured and were used as
a baseline in the nuclear modification factor for Pb+Pb collisions at
the same energy. Relative to $N_\mathrm{coll}$ scaling expected for
hard processes in the absence of nuclear effects a suppression of
$\pi^0$ yields for $p_\mathrm{T} \gtrsim 2$\,GeV/$c$ has been observed
in central ($N_\mathrm{part} \gtrsim 300$) Pb+Pb collisions.  This
observation is qualitatively consistent with models which assume
parton energy loss at $\sqrt{s_\mathrm{NN}} = 17.3$\,GeV.

\end{document}